\documentclass[reprint,aps,prl,showpacs,floatfix]{revtex4-1}
\usepackage{graphics,epsfig,graphicx}
\usepackage{amsbsy}
\usepackage{amsmath}
\usepackage{bm}
\usepackage{amsfonts}
\usepackage{cancel}
\usepackage{multirow}
\usepackage{slashed}
\begin{document}

\title{Correlations imposed by the unitary limit between few-nucleon systems,
nuclear matter and neutron stars}


\author{A. Kievsky$^1$, M. Viviani$^1$, D. Logoteta$^1$, I. Bombaci$^{1,2}$, and L. Girlanda$^{3,4}$} 
\affiliation{$^1$Istituto Nazionale di Fisica Nucleare, Largo Pontecorvo 3, 56127 Pisa, Italy}
\affiliation{$^2$Department of Physics, University of Pisa, 56127 Pisa, Italy}
\affiliation{$^3$Department of Mathematics and Physics, University of Salento, I-73100 Lecce, Italy}
\affiliation{$^4$Istituto Nazionale di Fisica Nucleare, Sezione di Lecce, I-73100 Lecce, Italy}

\begin{abstract}
The large values of the singlet and triplet two-nucleon scattering lengths locate
the nuclear system close to the unitary limit. This particular position strongly constrains
the low-energy observables in the three-nucleon system as depending on one parameter,
the triton binding energy, and introduces correlations in the low energy sector of light 
nuclei. Here we analyze the propagation of these correlations to infinite nuclear matter
showing that its saturation properties, the equation of state of $\beta$-stable 
nuclear matter and several properties of neutron stars, as their maximum mass, are well determined
solely by a few number of low-energy quantities of the two- and three-nucleon systems.
In this way we make a direct link between the universal behavior observed in the
low-energy region of few-nucleon systems and fundamental properties of nuclear matter and
neutron stars.
 \end{abstract}
\maketitle

{\sl Introduction.}  The unitary limit, characterized by the divergence of the $s$-wave 
two-body scattering length $a$, is a critical point in which the two-body system has no scale.
As the system approaches this limit, it presents a 
continuous scale invariance. The two-body scattering length appears as a control parameter: 
it determines the low-energy observables with a functional dependence dictated by dimensional analysis.
For identical particles the three-boson system presents a discrete scale symmetry,
governed by the size of a particular three-body state, and shows the
Efimov effect at the unitary point. 
The three-boson spectrum is then determined by the
control parameter $a$ and the three-body parameter $\kappa_*$, the binding momentum of the selected
state. All these features, collected in what is now called Efimov physics, are intensively studied 
from an experimental~\cite{ferlaino2011,machtey2012,roy2013,dyke2013} 
as well as a theoretical point of view (for recent reviews see Refs.~\cite{report,frederico2011}). 
The systems inside this window (Efimov window)
have many striking properties characterized by their insensitivity to the particular form of
the interaction, they show universal behavior. 

In atomic physics the study of Efimov physics is based on the experimental ability of tuning
the scattering length using Feshbach resonances. It is interesting to notice that
nuclear physics is naturally close to the unitary limit~\cite{koenig2017}. In fact, the deuteron
as well as the virtual $^1S_0$ state are very shallow two nucleon systems.
The energy scale from which their energy can be estimated is not 
directly related to $r_N$, the range of the nuclear force, but to the two-body scattering length 
in the corresponding spin channel:
$|E_S|\approx \hbar^2/m a_S^2$, with $S=0,1$ and $m$ the nucleon mass,
where the singlet and triplet scattering lengths, $a_0$ and $a_1$, are both large 
with respect to the typical length of the interaction, $r_N\approx1.4\;$fm.

Recent studies of nuclear systems as Efimov systems can be found in Refs.~\cite{koenig2017,
kievsky2016} up to $A=4$. It emerges that the triton, $^3$He and $^4$He are the lowest Efimov
states at the particular values of the ratio $r_S/a_S$, with $r_S$ the effective range.
The quantities $a_S$, $r_S$  appear as control parameters and finite size parameters
respectively whereas the binding momentum of $^3$H, $\kappa_T$, is the three-body
parameter. The binding energy of $^4$He, $B(^4{\rm He})=\hbar^2\kappa_\alpha^2/m$ 
can be deduced from the universal ratio
$\kappa_\alpha/\kappa_T\approx 1.9$, considering finite size and Coulomb corrections~\cite{koenig2017}.

The question that we want to discuss here is the constraints imposed by the location
of the nuclear system close to the unitary limit as they propagate with the number of particles.
In particular we will analyze the saturation properties of nuclear matter (NM) determined 
solely by $a_S$, $r_S$ and $\kappa_T$ and, more important, the equation of state (EoS)
of $\beta$-stable nuclear matter and the corresponding properties of neutron stars (NSs) and
particularly of their maximum mass configuration.
In this way we introduce a strict correlation between a few number of low-energy observables 
in $A=2,3,4$ systems and fundamental properties of NM and NS.

To follow this study we make use of the effective field theory (EFT) framework with and
without pions. In the latter case (pionless EFT) the leading order (LO) has been studied 
in a series of articles~\cite{bedaque1999,bedaque02} showing that it consists of two 
contact terms plus a contact three-body interaction needed to stabilize the 
three-nucleon system against the Thomas collapse. 
The pionless EFT is closely connected to the pioneering work by 
V. Efimov more than 40 years ago~\cite{efimov1,efimov2}. 
The LO parameters of the theory can be used to fix the low-energy parameters $a_S$ and $\kappa_T$
with the consequence that also the $^4$He binding energy is well reproduced
resulting in an energy per particle of about $7$ MeV. This quantity compares well with the 
average binding energy per particle along the nuclear chart of around $8$ MeV, having a peak of approximately 
$8.8$ MeV at the $^{56}{\rm Fe}$ nucleus. Accordingly the pionless LO maintains its character along 
the nuclear chart describing correctly the threshold at which nuclei bind beyond the clusterization 
on $\alpha's$. The binding beyond these thresholds may well be
considered as a higher order effect, that a LO description need not address in detail.
As an example we can mention the $\alpha+d$ threshold of $^6{\rm Li}$, the three- and 
four-$\alpha$ threshold of $^{12}{\rm C}$ and $^{16}{\rm O}$ and so on.   
Calculations using a pionless two- and three-body potentials beyond the LO will clarify further 
this point~\cite{kirscher2010,lensky2016,girlanda2011}. 

Correlations imposed by the unitary limit between few- and many-body systems have been discussed
in the context of EFT in Ref.~\cite{vankolck}.
Recent studies at LO have been done in light and medium-mass 
nuclei~\cite{stetcu,bansal,contesi} and boson systems~\cite{bira}.
Correlations between the triton and nuclear matter can be found in Ref.~\cite{delfino}. 
Clusters of bosons close to the unitary limit have been studied using a LO EFT-inspired potential
in Refs.~\cite{kievsky2011,gatto2011,gatto2012,kievsky2017b}.

{\sl The LO EFT-inspired potential.} In the following we define our LO EFT-inspired potential and fix the associate 
low energy constants from the low energy data in the two- and three-body
systems. The two-body potential we use, which includes all
LO interactions from EFT and some of its important finite range corrections is
\begin{equation}
V^{2N}_{LO}= V_{sr} +V_\pi
\end{equation}
where $V_{sr}$ is the short-range interaction and $V_\pi$ is the one-pion-exchange potential
(OPEP). The short-range interaction is a regularized contact interaction and has a spin dependence. It
can be written as
\begin{equation}
V_{sr}=C_0 V_0{\cal P}_{01} + C_1 V_1 {\cal P}_{10} \;\; ,
\end{equation}
where ${\cal P}_{ST}$ is a projector onto the total spin-isospin state $S,T$ of two nucleons.
Using a local gaussian regulator, the two potentials $V_0$ and $V_1$ have
the following form
\begin{equation}
V_S= V_S(r)= \frac{1}{\pi^{3/2}d_S^3} e^{-r^2/d_S^2}
\end{equation}
and $V_\pi$ is the regularized OPEP potential 
\begin{equation}
V_\pi(r)= {\bm \tau}_1\cdot{\bm \tau}_2
\left[ {\bm \sigma}_1\cdot{\bm \sigma}_2 Y_\beta(r)+ S_{12} T_\beta(r)\right] 
\end{equation}
with the central and tensor factors ($x=m_\pi r$)
\begin{equation}
\begin{gathered}
Y_\beta(x)= \frac{g_A^2 m_\pi^3}{12\pi F_\pi^2}\frac{e^{-x}}{x}(1-e^{-r^2/\beta^2)} \\
T_\beta(x)= \frac{g_A^2 m_\pi^3}{12\pi F_\pi^2}\frac{e^{-x}}{x}(1+\frac{3}{x}+\frac{3}{x^2})(1-e^{-r^2/\beta^2})^2\;\; .
\end{gathered}
\end{equation}
With this choice~\cite{kievsky2017}, the two-body potential 
$V^{2N}_{LO}$ has the form of the LO pionless EFT potential
($\beta\rightarrow\infty$) and tends continuosly to the LO potential in chiral perturbation theory.
The strength $C_S$ and range $d_S$ are designed to reproduce the $np$ scattering length 
and effective range, $a_S$ and $r_S$, in channels $S,T = 0,1$ and $1,0$ for different values of the
regulator $\beta$. Due to the shallow character of the deuteron
state and virtual ${}^1S_0$ state, this procedure automatically fixes the correct binding of these
two states. We now extend our analysis to the three- and four-nucleon systems. 
The effective potential is 
\begin{equation}
V_{LO}^{2N+3N}=\sum_{i<j}\left[V_{sr}(i,j)+V_\pi(i,j)\right] +\sum_{cyclic} W(i,j,k)
\end{equation}
where we have considered the possibility of a (regularized) contact three-body term of the form
\begin{equation}
W(i,j,k)=W_0 e^{-r_{ij}^2/r_3^2} e^{-r_{ik}^2/r_3^2} \;\; .
\label{eq:w30}
\end{equation}
We calculate the $^3$H and $^4$He energies, 
$B(^3{\rm H})$ and $B(^4{\rm He})$, for different values of $r_3$. In each case the strength
$W_0$ is fixed to reproduce $B(^3{\rm H})$. The results are shown in Fig.1 for different values of
the regulator of the OPEP $\beta$. As $\beta \rightarrow \infty$ the LO pionless
theory is recovered whilst the lowest value, $\beta=1.0\;$fm, is an extreme case, 
well inside the region $\beta<1/m_\pi\approx 1.5\;$fm corresponding to the formation of the OPEP 
tail. 
From the figure we observe that at low values of the range $r_3$
the curves are close to and slightly below the experimental binding of $28.3\;$ MeV whereas
for $r_3 > 3\;$fm the curve tend to be above $30$ MeV. The $\beta=1\;$fm  curve remains very stable
and limit the two regions from above (for the lowest $r_3$ values) and from below (for the
highest $r_3$ values).

\begin{figure}
\includegraphics[height=\linewidth,angle=-90]{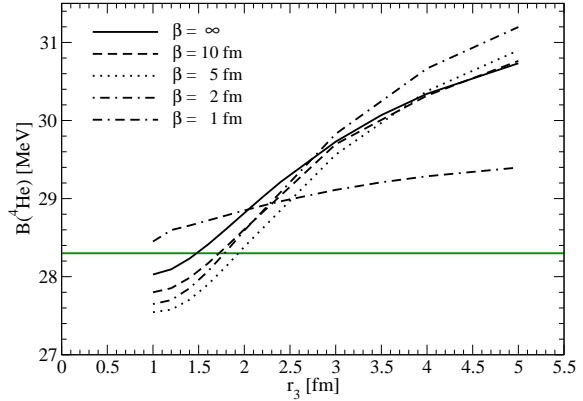}
\caption{$B(^4{\rm He})$ as a function of the three-body range $r_3$
for different values of the OPEP regulator $\beta$.}
\end{figure}

The analysis with equal values of the two-body and three-body ranges, $d_0=d_1=r_3$,
has been done in Ref.~\cite{kievsky2017}. Here we vary the three-body range $r_3$ independently 
of the two-body ranges, fixed in the two-body sector by the effective range values. 
The interest of the present study is to analyze the impact of the low-energy properties, 
$a_S$, $r_S$, $B(^3{\rm H})$ and $B(^4{\rm He})$ in the determination of NM properties. 
Taking fixed the two-body parameters and $B(^3{\rm H})$, the three-body range $r_3$ allows
for a simultaneous description of $B(^3{\rm H})$ and $B(^4{\rm He})$.
It helps to construct 
a curve of the energy per particle of symmetric nuclear matter with a saturation point
as close as possible to the empirical one. Specifically at $\beta \rightarrow \infty$ the two-body
ranges are $d_0=1.83\;$fm, $d_1=1.56\;$fm and $B(^4{\rm He})$ is well reproduced with $r_3=1.5\;$fm.
Instead at $\beta =2\;$ fm, $d_0=1.54\;$fm, $d_1=1.39\;$fm and $r_3=1.7\;$fm. 

{\sl Nuclear matter}.
We next discuss the application of the two- and three-nucleon forces, derived in the 
previous section, to the case of NM.     
To calculate the energy per nucleon $E/A$ of NM we make use of the the 
Brueckner--Bethe--Goldstone (BBG) quantum many-body theory (see e.g.~\cite{bbg1,bbg2} and references 
therein) considering contributions up to two-hole-line level, 
the so called Brueckner--Hartree-Fock (BHF) approximation.  
The BHF approximation incorporates in an exact way the two particle correlations via a self-consistent  
determination of the $G$ matrix and the single-particle auxiliary potential $U(k)$, for which   
we use the continuous choice \cite{jeuk+67,baldo+90}. 
As shown in \cite{song98,baldo00}, the contribution of the three-hole-line diagrams
is minimized in this prescription indicating a fast convergence of  
the hole-line expansion for $E/A$.  
In our calculations the three-nucleon force has been reduced to an effective density dependent 
two-body force by averaging over the coordinates (momentum, spin and isospin) of one of the nucleons 
as described in Ref.~\cite{LBK2}.   

\begin{figure}
\includegraphics[width=9cm,height=10cm,angle=-90]{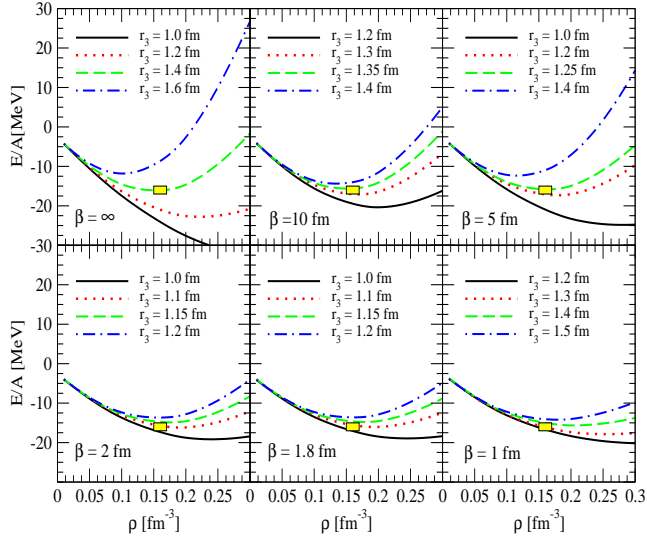} 
\caption{Energy per particle of SNM $E/A$ as a function of the nucleonic 
density $\rho$ for several combinations of the two- and three-body interactions.}
\label{fig2}
\end{figure}

The energy per particle $E/A$ of symmetric nuclear matter (SNM) is shown in Fig.\ \ref{fig2} 
for various parametrizations of the two- and three-body forces. In each panel, for a fixed value 
of the OPEP regulator $\beta$ of the two-body force, we show the saturation curve 
(i.e. $E/A$ as a function of the nucleonic density $\rho$) of SNM obtained using four different 
values of the three-nucleon force range $r_3$.  
The empirical saturation point of SNM ($\rho_{0} = 0.16 \pm  0.01~{\rm fm}^{-3}$, 
$E/A|_{\rho_0} = -16.0 \pm 1.0~{\rm MeV}$) is denoted by a gray box in each panel 
of Fig.\ \ref{fig2}.

\begin{table}[h]
\begin{tabular}{ccccccc}
\hline
$\beta$ & $r_3$ & $\rho_0$    & $E/A|_{\rho_0}$ &$E_{sym}^0$& $L$  & $K_\infty$  \\
 (fm)   &  (fm) & (fm$^{-3}$) &           (MeV) &  (MeV)    & (MeV)& (MeV)       \\
\hline
$\infty$ &  1.4   &    0.151   &   -16.11   &    35.20     &      70.2       &     251 \\
  10     &  1.35  &    0.150   &   -15.65   &    34.92     &      69.8       &     251 \\
   5     &  1.25  &    0.160   &   -15.80   &    36.16     &      71.0       &     247 \\
   2     &  1.15  &    0.173   &   -14.83   &    36.37     &      67.9       &     209 \\
   1.8   &  1.15  &    0.176   &   -14.74   &    36.33     &      67.0       &     205 \\
   1     &  1.5   &    0.179   &   -14.20   &    35.02     &      58.5       &     203 \\
\hline
\end{tabular}
\caption{Nuclear matter properties at the calculated saturation density $\rho_0$ (3$^{rd}$ column)
for different combinations of the interaction model parameters $\beta$ and $r_3$.
$E/A|_{\rho_0}$ is the SNM saturation energy, $E_{sym}^0$ the symmetry energy,  $L$ the symmetry energy slope parameter and $K_\infty$ the SNM incompressibility. }
\vspace{1cm}
\label{tab1}
\end{table}

The calculated saturation points for the "best" saturation curve in each panel of 
Fig.\ \ref{fig2} are reported in Tab.\ \ref{tab1}. 
We note that the empirical saturation point of SNM is adequately reproduced by the first three entries, 
($\beta = \infty$, $r_3 = 1.4\;$fm), ($\beta = 10\;$fm, $r_3 = 1.35\;$fm) and  
($\beta = 5\;$fm, $r_3=1.25\;$fm), in Tab.\ \ref{tab1}.  
For smaller values of $\beta$, the empirical saturation point of SNM can not be reproduced. 
However, even in this case, optimizing the value of the parameter $r_3$, a 
reasonable saturation point is obtained (see last three entries in Tab.\ \ref{tab1}). 

\begin{figure}   
\includegraphics[width=7.5cm,angle=-90]{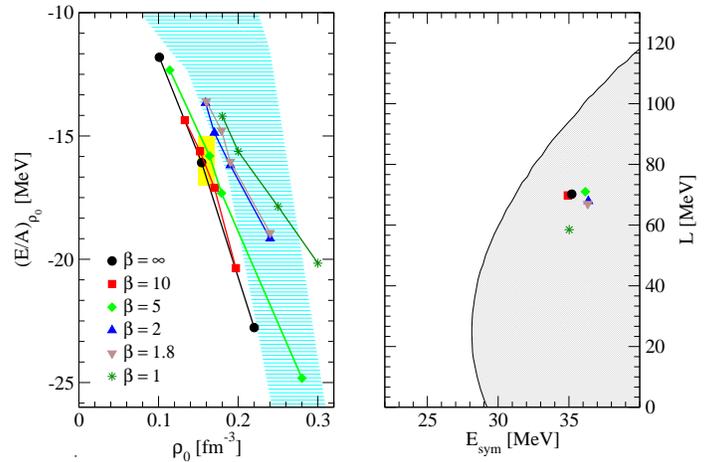} 
\caption{Left panel: saturation points of SNM for the various interaction models considered in this work. 
Different symbols represent the saturation point for different values of the OPEP regulator 
$\beta$ of the two-body force. 
Points with the same symbol represent the saturation point for different values of the 
three-nucleon force range $r_3$ (the value of $r_3$ increases from the bottom to the top of each line). 
For the explanation of the hatched zone see text. 
Right panel: Calculated values (points with different symbols) of $E_{sym}^0$ and $L$ for the six interaction models reported in Tab.\ \ref{tab1}.   
The gray zone represents the allowed region in the $E_{sym}^0$--$L$ plane which is 
compatible with the unitary gas bound proposed by the authors of Ref.\ \cite{tews2017}. }
\label{fig3}
\end{figure}

The calculated saturation points of SNM for various interaction models are shown in the left panel 
of Fig.\ \ref{fig3}. Here the empirical saturation point is denoted by a yellow box.  
Notice that for fixed $\beta$, the saturation points of the various interaction models show an 
almost linear dependence on the three-nucleon force range $r_3$ (the value of $r_3$ increases 
from the bottom to the top of each line). 
All together the calculated saturation points locate a narrow band, the so called Coester 
band\ \cite{coester70,day81}, which for the interaction models used goes through the empirical 
saturation point. 
The hatched zone in this panel collects calculations in which the two-body and 
three-body ranges have been taken equal $d_0=d_1=r_3$. In this case
the two-body scattering lengths, $a_S$, and $B(^3{\rm H})$ 
have been kept fixed by proper values of the strengths $C_S$ and $W_0$. 
These results give rise to a much broader Coester band, with saturation points that vary in a much
larger range and do not overlap with the empirical saturation point of SNM 
contrary to the calculations in which only the three-body range $r_3$ is varied.
These results show the importance of tuning the $r_3$ parameter.

The nuclear symmetry energy $E_{sym}$, and particularly its density dependence, is an important 
physical quantity which regulates the properties of asymmetric NM (i.e. matter with 
$\rho_n \neq \rho_p$, with $\rho_n$ and $\rho_p$ being the neutron and proton densities respectively).   
The symmetry energy can be obtained \cite{bl91} taking the difference between the energy per 
nucleon $E/A$ of pure neutron matter and the one of SNM at a given total nucleon number density 
$\rho = \rho_n + \rho_p$.  
The symmetry properties of NM around the saturation density $\rho_0$ are summarized 
by the value of $E_{sym}^0 \equiv E_{sym}(\rho_0)$ and by the value of the 
so called symmetry energy slope parameter 
\begin{equation}
       L = 3 \rho_{0} \frac{\partial E_{sym}(\rho)}{\partial \rho}\Big|_{\rho_{0}} 
\label{slope}
\end{equation} 

It has been shown \cite{latt2013} that a strong correlation between the values of   
$E_{sym}^0$ and $L$ can be deduced in a nearly model-independent way from nuclear binding energies.   
In addition, it has been recently demonstrated \cite{tews2017} that the unitary gas limit \cite{zwier2015}, 
which can be used to describe low density neutron matter, puts stringent constraints on the possible 
values of the symmetry energy parameters, excluding an ample region in the $E_{sym}^0$--$L$ plane 
(the white region on the left of the line in the right panel of Fig.\ \ref{fig3}).  
As pointed out by the authors of Ref.~\cite{tews2017} several EOS models currently used in   
astrophysical simulations of supernova explosions and binary neutron star mergers violate the 
unitary gas bounds. Thus the unitary gas model can be used as a novel way to constrain dense matter EOS 
in astrophysical applications~\cite{endrizzi}.    

The values of $E_{sym}^0$ and $L$ calculated for our "best" interaction models are reported 
in Tab.\ \ref{tab1} and are plotted in the right panel of Fig.\ \ref{fig3}. 
As one can see our calculated $E_{sym}^0$ and $L$ are totally compatible with the unitary gas 
bound (the gray zone in the right panel of Fig.\ \ref{fig3}) proposed in Ref.\ \cite{tews2017}.  

Next, in the last column of Tab.~\ref{tab1},  we report the incompressibility of SNM   
\begin{equation}
       K_\infty = 9 \rho_{0}^2 \, \frac{\partial^2 E/A}{\partial \rho^2}\Big|_{\rho_{0}}  
\label{incom}
\end{equation}
at the calculated saturation point for each of the interaction models listed in Tab.\ \ref{tab1}.    
Our calculated values for $K_\infty$ are in very good agreement with the empirical value    
$K_\infty = 210 \pm 30$ MeV \citep{blaizot76} or more recently $K_\infty = 240 \pm 20$~MeV \citep{shlo06}   
extracted from experimental data of giant monopole resonance energies in medium-mass and heavy nuclei. 

Finally for the three interaction models which reproduce the empirical saturation point of SNM 
(first three entries in Tab.\ \ref{tab1}) we have calculated the EoS of $\beta$-stable 
NM (see e.g. \cite{prak97,BL2018}) and then integrated the stellar structure equations in 
general relativity for non rotating stars. The results of our calculations for the stellar maximum mass 
configuration are reported in Tab.\ \ref{tab_NS}.   
It should be noticed that the neutron star matter EOS for the interaction models listed in Tab.\ \ref{tab_NS} 
are all compatible with present measured NM masses and particularly with the mass   
$M = 2.01 \pm 0.04 \, M_{\odot}$ \cite{anto13} of the NS in PSR~J0348+0432
and 
$M = 2.27^{+0.17}_{-0.15} \, M_{\odot}$ \cite{linares} of the NS in PSR~J2215+5135.

\begin{table}[h] 
\caption{Neutron star properties for the maximum mass configuration for the interaction 
parameters reported in the first two columns. $M_{max}$ is the stellar gravitational maximum mass 
(in unit of the mass of the Sun $M_\odot = 1.989\times 10^{33}$~g), $R$ is the corresponding radius 
and $\rho_c$ the central nucleonic density.} 
\label{tab_NS}
\small
\centering
\begin{tabular}{ccccc}
\hline
$\beta$ (fm) & $r_3$ (fm) & $M_{max}$ ($M_\odot$) & $R$ (km) & $\rho_c$ (fm$^{-3}$)  \\
\hline 
$\infty$  &    1.40    &   2.52  &   11.64  & 0.84 \\   
   10     &    1.35    &   2.52  &   11.68  & 0.82 \\   
    5     &    1.25    &   2.46  &   11.29  & 0.89 \\   
\hline
\end{tabular}
\end{table} 

{\sl Conclusions}.
We have analyzed correlations between observables in the few-nucleon sector, 
NM and NSs caused by the location of the nuclear system close to the 
unitary limit. The LO EFT-inspired potential, based on pionless
EFT plus a regularized OPEP term, has been constructed to describe two-body low-energy
observables and the triton binding energy.
The three-body range, $r_3$, was allowed to vary and,
in terms of this quantity and for different regularizations of the OPEP,
we have calculated $B(^4{\rm He})$, 
the energy per particle of SNM (and the corresponding saturation point), 
the nuclear symmetry energy, its slope parameter, the SNM incompressibility
and the EoS of $\beta$-stable NM.  
Unexpectedly this very simple potential, in particular for values making the
OPEP small, reproduces many of the mentioned observables.
Moreover the EoS of $\beta$-stable matter produces neutron star configurations with a 
maximum mass compatible with present measured neutron 
star masses \cite{anto13,linares}.  
This analysis indicates that the unitary limit, which controls the universal aspects of the two-body
physics, introduces severe constraints in the nuclear system taking priority over the precise
description of the two-nucleon data up to high energies.


Due to the simplicity of the model,
the values of $r_3$ at which these results are obtained are slightly smaller than the
best values needed for describing $B(^4{\rm He})$. The feature that the requested 
three-body range be shorter than the two-body ranges can be understood
from the way the contact interactions are regularized: if the momentum transfers 
${\bf k_1}$ and ${\bf k_2}$ of two particles are limited to a range $\Lambda$, the third one, 
${\bf k_3}$, constrained by momentum conservation to be the sum of the two, may take larger values; 
with gaussian cutoffs its width would be a factor $\sqrt{2}$ larger, implying the necessity of 
shorter-range ranges for the regularized three-body contact interaction.

The main result of this study is to put in evidence the direct connection between
many-body observables and the deuteron and the $S=0$ virtual state scales given by
$a_S$ and the triton binding energy whose value fixes the strength of the three-body potential
$W_0$. This result extends the analysis of Refs.~\cite{koenig2017,kievsky2017} to
infinite nuclear systems showing that fundamental many-body properties are controlled
by the position of the nuclear system close to the unitary limit.


\end{document}